\documentclass[prd,preprint,amsmath,amssymb]{revtex4}%
\usepackage[dvips]{graphics}
\usepackage{array}

\newcommand{\T}{{\cal T}}

\begin{document}

\title{Color confinement in Coulomb gauge QCD}

\author{A.~Nakamura }
\affiliation{Research Institute for Information Science and Education,
 Hiroshima University, Higashi-Hiroshima 739-8521, Japan}
\author{T.~Saito}
\affiliation{Research Institute for Information Science and Education,
 Hiroshima University, Higashi-Hiroshima 739-8521, Japan}

%\author{A.~Nakamura and T.~Saito}
%\inst{Research Institute for Information Science and Education,
% Hiroshima University, Higashi-Hiroshima 739-8521, Japan}

\begin{abstract}
%\abst{
We study the long-range behavior of the heavy quark potential
in Coulomb gauge using a quenched $SU(3)$ lattice gauge simulation
with partial-length Polyakov line correlators.
We show that the Coulomb heavy quark potential associated with
 the instantaneous part of gluon propagators
 in Coulomb gauge, presents a linearly rising behavior at large distances,
 and the resulting Coulomb string tension is greater
 than the Wilson loop string tension,
 which can be explained by Zwanziger's inequality.
The linearly rising behavior of the Coulomb heavy quark potential 
persists even in the deconfinement phase.
The heavy quark potential in Lorentz gauge shows completely different
 behavior than that in Coulomb gauge.
Our $SU(3)$ result, i.e.,  the Coulomb heavy quark potential is confining, 
qualitatively agrees with that of the $SU(2)$ analysis 
carried out by Greensite, Olejnik and Zwanziger.
%}
\end{abstract}

%\end{frontmatter}

%\pacs{12.39.Mk, 12.38.Aw, 12.38.Gc, 11.15.Ha}
%\keywords{lattice QCD, singlet, Coulomb confinement}

\maketitle

\section{Introduction}
Recently, the color confinement scenario in Coulomb gauge has been 
revealing its importance 
\cite{CCG,PRC,renorm,Zwan,minimal,Greensite,Greensite2,Reinhardt}.
Zwanziger \cite{CCG} discussed the significance of a color-Coulomb
potential in color confinement.
He and his collaborators showed that, in Coulomb gauge, 
the time-time component of gluon propagators, $g^2 D_{00}$, 
including the instantaneous color-Coulomb potential and
the non-instantaneous vacuum polarization, 
is invariant under renormalization \cite{CCG,PRC},
where $g$ is a coupling constant of gauge field theory.
The instantaneous color-Coulomb potential plays an essential role
in the Coulomb gauge confinement scenario.
In the $SU(2)$ numerical simulation carried out 
by Cucchieri and Zwanziger \cite{minimal}, 
it was found that
the instantaneous color-Coulomb potential $D_{00}(\vec{k})$
is strongly enhanced at $\vec{k}=0$.
Moreover, Zwanziger pointed out that there exists the inequality\cite{Zwan} 
\begin{equation}
V_{phys}(R) \le V_{coul}(R) ,
\label{ZwanzigerIneq}
\end{equation}
where $V_{phys}(R)$ means the physical heavy quark potential
and $V_{coul}(R)$ the Coulomb heavy quark potential 
 corresponding to the instantaneous part of $D_{00}$.
This inequality indicates that
if the physical heavy quark potential is confining,
then the Coulomb heavy quark potential is also confining.
Furthermore, in $SU(2)$ lattice simulations
Greensite et al. found that the Coulomb heavy quark potential
grows linearly at large quark separations \cite{Greensite,Greensite2}. 
They showed that
the instantaneous part of $D_{00}$ can be nonperturbatively
managed with a partial-length Polyakov line (PPL) correlator. 
(See Ref. \cite{DZ-ptps98} for an excellent review.)

The Coulomb gauge fixing does {\it not} fix a gauge completely
since it leaves the temporal-gauge field free; i.e., 
one may consider that Coulomb gauge has a remnant residual symmetry 
in the temporal direction.
Marinari et al. \cite{Marinari} conjectured that 
the Coulomb remnant symmetry breaking may result in
a new order parameter for the deconfinement phase transition.
Recently, $SU(2)$ gauge-Higgs theory was investigated numerically
from this point of view \cite{Greensite2}.

In this work,
we study the heavy quark potential in 
the color-singlet channel nonperturbatively,
 using quenched $SU(3)$ lattice QCD simulations
with the PPL correlator in Coulomb gauge.
We first investigate the behavior of the Coulomb heavy quark potential 
in the confinement and deconfinement phases.
The most interesting point is
whether the Coulomb heavy quark potential has
a linearly rising feature at large quark separations.
We confirm the consistency
between the Coulomb heavy quark potential and
the usual Wilson loop potential (or the Polyakov line potential).
We repeat the same calculation in Lorentz gauge, 
and compare the results obtained in the two cases.
We examine the Coulomb remnant symmetry breaking
at zero and finite temperatures.

This paper is organized as follows:
In section II, we briefly present definitions of the PPL correlators
and the Coulomb heavy quark potential in the color-singlet channel.
We next summarize the gauge fixing methods and 
the order parameter for the Coulomb remnant symmetry.
In section III, the numerical results are shown.
Concluding remarks are given in section IV.

\section{Partial length Polyakov line}
\label{PPL}

In this section, we introduce 
the Coulomb potential in the singlet channel between two static heavy quarks 
and describe how to fix a gauge on the lattice
and the order parameter for the Coulomb remnant symmetry.

Partial length Polyakov lines (PPL) are defined as
\cite{Greensite,Greensite2}
\begin{equation}
L(\vec{x},T) = \displaystyle\prod_{t=1}^{T} U_0(\vec{x},t),
\quad T=1, 2, \cdots, N_t.
\end{equation}
Here $U_0(\vec{x},t)= \exp(iagA_0(\vec{x},t))$
is an $SU(3)$ link variable in the temporal direction and 
$a$, $g$, $A_0(\vec{x},t)$ and $N_t$ represent the lattice cutoff,
the gauge coupling, the time component of the gauge potential and
the temporal lattice size. 
The PPL correlators in the color-$SU(3)$ singlet channel are given by
\begin{equation}
G(R,T) = \frac{1}{3}\left< Tr[L(R,T)L^{\dagger}(0,T)] \right>,\label{cor}
\end{equation}
where $R = \arrowvert \vec{x} \arrowvert $.
From these correlators, 
we can evaluate the color-singlet potentials on the lattice, 
\begin{equation}
V(R,T) = \log \left[
\frac{G(R,T)}{G(R,T+a)} \right]\label{pot1}. 
\end{equation}
For the smallest temporal lattice extension, i.e., $T=0$,
we define
\begin{equation}
V(R,0) = - \log [ G(R,a) ]\label{pot2}.
\end{equation}

The potential $V(R,0)$ in Coulomb gauge corresponds to
 a color-Coulomb potential,
$V_{coul}(R)$;
 the time-time component of the gluon propagator,  
$D_{00}(\vec{x},t)=\langle A_0(\vec{x},t)A_0(\vec{0},t)\rangle$, 
can be decomposed into the non-instantaneous vacuum polarization, 
$P(\vec{x},t)$, 
plus the instantaneous color-Coulomb potential, $V_{coul}(R)\delta(t)$, 
whose term dominates in the Coulomb confinement scenario \cite{renorm}:
\begin{equation}
D_{00}(\vec{x},t) =
V_{coul}(R)\delta(t) + P(\vec{x},t).
\end{equation}
In the numerical simulations, the instantaneous contribution
has been managed through $V(R,0)$ \cite{Greensite,Greensite2}; 
this also appears as the enhancement of $D_{00}$
at vanishing momentum \cite{minimal}.
In the limit $T \rightarrow \infty$, 
 $V(R,T)$ further corresponds to 
%The $V(R,T)$ at $T \rightarrow \infty$ further corresponds to 
a physical potential, $V_{phys}$, which can usually be regarded as   
the Wilson loop potential in the same limit.
In addition, these two potentials satisfy Zwanziger's inequality, 
Eq. (\ref{ZwanzigerIneq}) \cite{Zwan}; i.e., 
if color confinement exists, then the color-Coulomb potential
is also confining.

We use the Coulomb gauge realized on the lattice as 
\begin{equation}
\mbox{Max} \sum_{\vec{x}} \sum_{i=1}^3 \mbox{ReTr} U_i^{\dagger}(\vec{x},t)  
\end{equation}
by repeating the gauge rotations 
\begin{equation}
U_i(\vec{x},t) \rightarrow U_i^{\omega}(\vec{x},t)
= \omega^{\dagger}(\vec{x},t)U_i(\vec{x},t) 
\omega(\vec{x}+\hat{i},t),
\end{equation}
where 
\footnote{In this study, we adopt
 $\omega = e^{i \alpha \partial_i A_i}$ as the gauge rotation matrix, and 
the parameter $\alpha$ is chosen suitably,
 depending on the lattice size, etc.
}
{$\omega$ $\in SU(3)$} is a gauge rotation matrix and 
$U_i(\vec{x},t)$ are spatial lattice link variables.
\footnote{Here we did not investigate the Gribov copy effect.}
{Thus, each lattice configuration can be gauge fixed iteratively
\cite{Mandula}.}

The temporal gauge fields possess gauge freedom even after 
the Coulomb gauge fixed.
One can still perform a time dependent gauge rotation
on the Coulomb gauge fixed links: 
\begin{equation}
\begin{array}{ccl}
 U_i(\vec{x},t) &\rightarrow&
 \omega^{\dagger}(t) U_i(\vec{x},t) \omega(t), \\
 U_0(\vec{x},t) &\rightarrow&
 \omega^{\dagger}(t) U_0(\vec{x},t) \omega(t+1) \label{gt}.
\end{array}
\end{equation}
It was conjectured by Marinari et al. \cite{Marinari}
that this remnant symmetry in Coulomb gauge is closely related
to color confinement physics.
Recently, Greensite et al. \cite{Greensite2}
proposed the following order parameter for the Coulomb remnant symmetry: 
\begin{equation}
Q_s = \frac{1}{N_t}
\sum_{t=1}^{N_t} \left<\sqrt{\frac{1}{3}
\mbox{Tr} 
\left[ \tilde{U}(t) \tilde{U}^{\dagger}(t) \right] 
}\right>, \label{qs}
\end{equation}
\begin{equation}
\tilde{U}(t) = \frac{1}{V_s} \sum_{\vec{x}} U_0(\vec{x},t),
\end{equation}
where $V_s=N_s^3$, and $N_s$ stands for the spatial lattice size.
The quantity $Q_s$ is invariant under the transformation (\ref{gt}).
Therefore $Q_s=0$ if the symmetry is not broken.
This quantity contains
\begin{equation}
\frac{1}{V_s^2} \sum_{\vec{x},\vec{y}} 
\mbox{Tr} U_0(\vec{x},t)U_0(\vec{y},t)
\sim
\frac{1}{V_s^2} \sum_{\vec{x},\vec{y}} e^{-V(|\vec{x}-\vec{y}|)} . 
\end{equation}
Consequently, if $V$ increases linearly for large $|\vec{x}-\vec{y}|$, i.e., 
 it is a confining potential, then $Q_s \rightarrow 0$ as 
$V_s \rightarrow \infty$, and if $V \rightarrow \mbox{const.}$ as 
$|\vec{x}-\vec{y}| \rightarrow \infty$, then
$Q_s > 0$.  
Hence, we may consider that $Q_s$ is an order parameter of the
confinement and deconfinement phase transitions. 
If this is the case, it would be a very desirable order parameter
because this works also for the full QCD including dynamical quarks.

\section{Simulation results}

In this study, we carried out $SU(3)$ lattice gauge simulations 
in the quench approximation to calculate the PPL correlators 
in the color-singlet channel.
The lattice configurations were generated by 
the heat-bath Monte Carlo technique
with a plaquette Wilson gauge action, and
we adopted the iterative method \cite{Mandula} for fixing a gauge.

\subsection{Coulomb heavy quark potential}

\begin{figure}[htbp]
\begin{center}
\resizebox{13cm}{!}{ \includegraphics{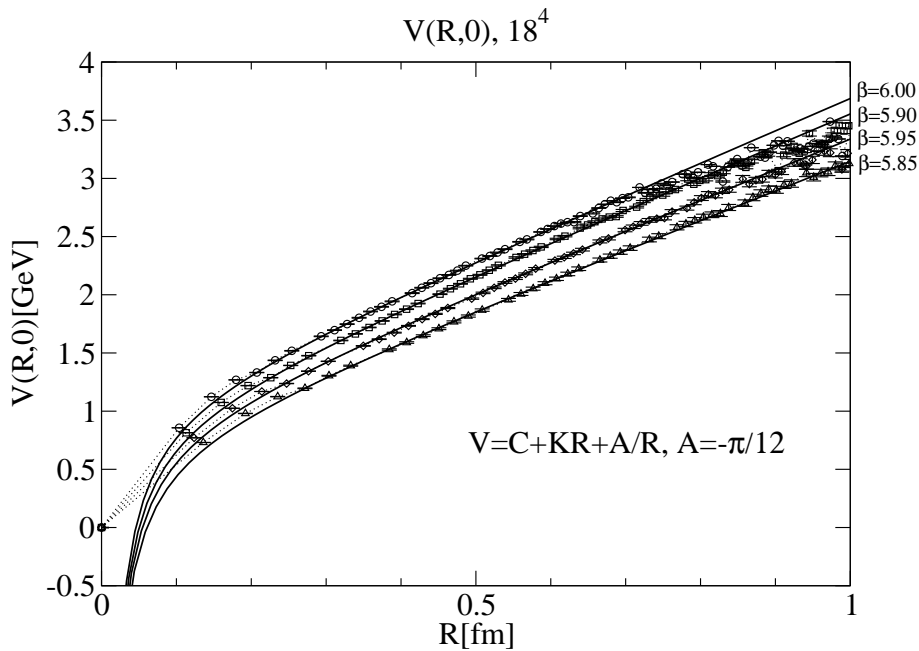} }
\caption{Coulomb heavy quark potentials
obtained from the PPL correlator with $T=1$.
These data were obtained
in the $18^4$ lattice simulation at $\beta = 5.85 - 6.00$.
We used 400 gauge configurations measured  every 100 sweeps.
}\label{sv}
\end{center}
\end{figure}

\begin{table}[h]
\caption{The fitting results for the string tensions.
$K_1$ in lattice units and $\sigma_1$ in physical units
were obtained from the singlet V(R,0). 
$K_w$ and $\sigma_w$ stand for the Wilson loop string tension at $\beta=6.0$ 
\cite{Bali} (not calculated here).
We use the relation $\sqrt{\sigma}=\sqrt{K}a^{-1}$ and 
the lattice cutoffs estimated with the Monte Carlo renormalization 
analyses \cite{qcdtaro}. }
\begin{center}
\begin{tabular}{lllll}
\hline
\hline
$\beta$ & $K_1$ & $\sqrt{\sigma_1}$ [MeV]& $K_w$ & $\sqrt{\sigma_w}$ [MeV] \\
\hline
5.85 & 0.2291(22) & 706(4)& &\\
5.90 & 0.1950(10) & 716(4)& &\\
5.95 & 0.1726(6)  & 736(3)& &\\
6.00 & 0.1467(4)  & 740(3)& 0.0513(25) \cite{Bali} & 470(46)\\
\hline \label{tab1}
\end{tabular}
\end{center}
\end{table}

Figure \ref{sv} shows the results of
 the Coulomb heavy quark potential $V(R,0)$ 
obtained from the PPL correlator with $T=1$.
These data were obtained from the $18^4$ lattice simulations
 at $\beta = 5.85 - 6.00$.
To obtain the string tension, we assume the following fitting function:
\begin{equation}
V(R,T) = C + KR + A/R,\quad A = - \pi/12, \label{fitting}
\end{equation}
where $C$ is a constant and $K$ corresponds to the string tension.
We find that the fittings are good; $\chi^2/ndf \sim O(1)$
 for the fitting range $R=2-6$.
It is found that the Coulomb heavy quark potential $V(R,0)$ rises linearly
as the distance $R$ increases at $\beta = 5.85 - 6.00$, and hence
it can be described by the linear rise function with the string tension.
The string tension for $\beta = 5.85 - 6.00$ and the Wilson loop 
string tension \cite{Bali} for $\beta=6.0$ are listed in Table \ref{tab1}.
$K_1$ at $\beta=6.0$ is
approximately three times larger than $K_w$,  
and the value of $\sqrt{\sigma_1}$ increases with $\beta$.
Although similar results for the $\beta$ dependence of the string tension 
were also obtained in the $SU(2)$ lattice 
simulations \cite{Greensite,Greensite2}, we have no clear explanation. 

\begin{figure}[htbp]
\vspace{0.1cm}
\begin{center}
 \resizebox{13cm}{!}{ \includegraphics{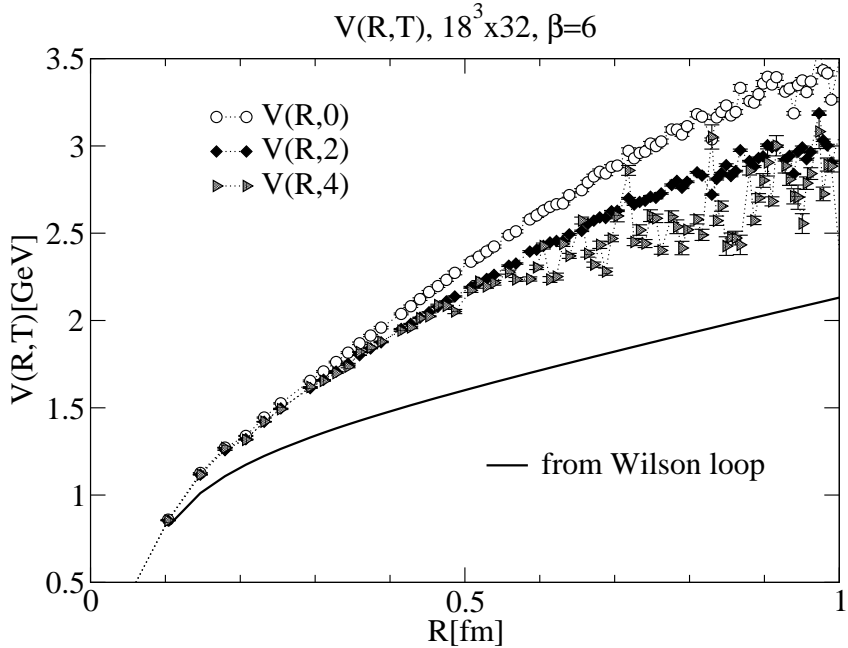} }
\caption{$T$ dependence of the Coulomb heavy quark potential
at $\beta = 6.0$.
The symbols with error bars and the solid curve correspond to 
the numerical data calculated in the present work  
and using the Wilson-loop potential \cite{Bali},
respectively.
As $T$ increases, $V(R,T)$ seems to approach $V_w(R)$. 
}\label{tdep}
\end{center}
\end{figure}

Because $V(R,T)$ in the limit $T \rightarrow \infty$ corresponds to a 
physical potential,
we expect that $V(R,T)$ becomes comparable with
the Wilson loop potential, $V_w(R)$, when $T$ becomes large enough.
The $T$ dependence of the Coulomb heavy quark potential
at $\beta=6.0$ is displayed in Fig. \ref{tdep}.
We used the $18^3 \times 32$ lattice and 600 configurations measured 
every 100 sweeps.
 $V(R,T)$ may approach $V_w(R)$ as $T$ increases.
\footnote{In this calculation,  
 the error bars at large distances 
estimated by the jackknife method seem to be relatively small. 
However, as the distance increases, 
 the lattice calculation of the PPL correlators 
becomes difficult, conceivably due to the smallness of
 $\langle \mbox{Tr} L(x,T) \rangle$ at large $T$.
To obtain more reliable data at large distances, 
the larger lattices and a smearing method are required.
}
{The $T$ dependence of the PPL correlator is not controlled completely here 
although there is consistency in the $SU(2)$ lattice calculations
in Refs. \cite{Greensite} and \cite{Greensite2}.}

\begin{figure}[htbp]
\begin{center}
\resizebox{13cm}{!}{ \includegraphics{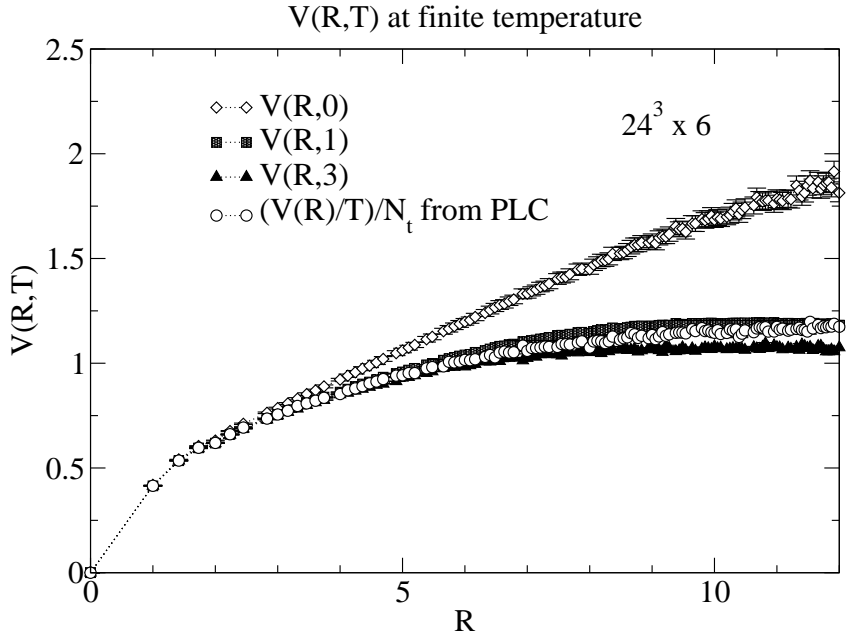} }
\caption{Linear-rise and screened potentials
in the deconfinement phase at $\T/\T_c \sim 1.50$.
Note that $V(R,0)$ still indicates linearly confining behavior,
 while the potentials $V(R,T)$
 with finite $T$ are screened and are consistent with the 
calculation employing the usual full Polyakov-line correlator.
}\label{spd}
\end{center}
\end{figure}

The finite temperature behavior of the Coulomb heavy quark potential
is shown in Fig. \ref{spd}. This simulation was carried out 
on the $24^3 \times 6$ lattice at $\beta=6.11$ ($a \sim 0.086$ fm),
corresponding to $\T/\T_c \sim 1.50$, where 
$\T$ stands for the system temperature and 
$\T_c$ the critical temperature of the quark-gluon plasma phase transition.
The 300 configurations measured every 100 steps are used.
It is very remarkable that even 
in the deconfinement region, $\T/\T_c \sim 1.50$, 
the Coulomb heavy quark potential $V(R,0)$ is {\it not} screened and 
still a linearly increasing function at large distances.
Fitting these data with Eq. (\ref{fitting}), 
 we obtain $K_1(\beta=6.11, \T/\T_c \sim 1.50)$ = 0.118(1),
 and we find that $\sqrt{\sigma_1}(\beta=6.11, \T/\T_c \sim 1.50)$
= 792(10)MeV, which is larger than
the value $\sqrt{\sigma_1}(\beta=6.0, \T \sim 0)$ = 740(4) MeV in 
Table \ref{tab1}; thus, it seems that the Coulomb string tensions
with $T=1$ depend on QCD coupling $\beta$ rather than
 the system temperature $\T$.
In the case of the color $SU(2)$ simulation, it is reported in Ref.
\cite{Greensite2} that the Coulomb string tension scales well
according to the two-loop $\beta$ function.
On the other hand, as the temporal extension $T$ increases,
the potential $V(R,T)$ are screened at large distances, 
$ R \gtrsim 1/\T=6$ on this lattice, and
they show the usual finite-temperature screening dynamics
that have been investigated nonperturbatively in lattice simulations 
\cite{CDP1,SGluon}.

\subsection{Remnant symmetry in Coulomb gauge}

\begin{figure}[htbp]
\begin{center}
\resizebox{13cm}{!}{\includegraphics{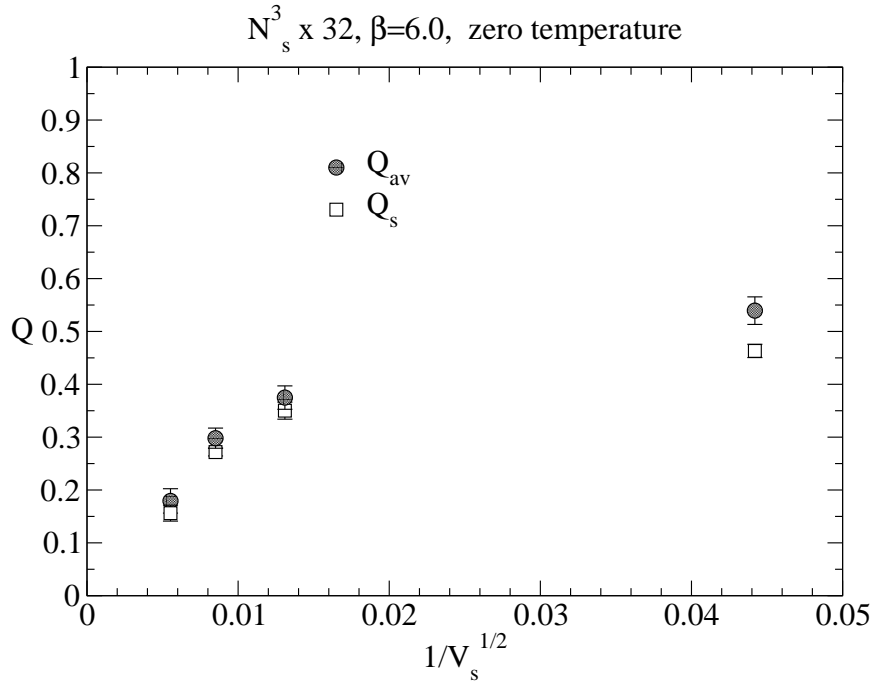}}
\caption{
Volume dependence of $Q$ in the confinement phase at zero temperature.
$Q_s$ and $Q_{av}$ vanish as $V \rightarrow \infty $.
}\label{sqv}
\end{center}
\end{figure}

\begin{figure}[htbp]
\begin{center}
\resizebox{13cm}{!}{\includegraphics{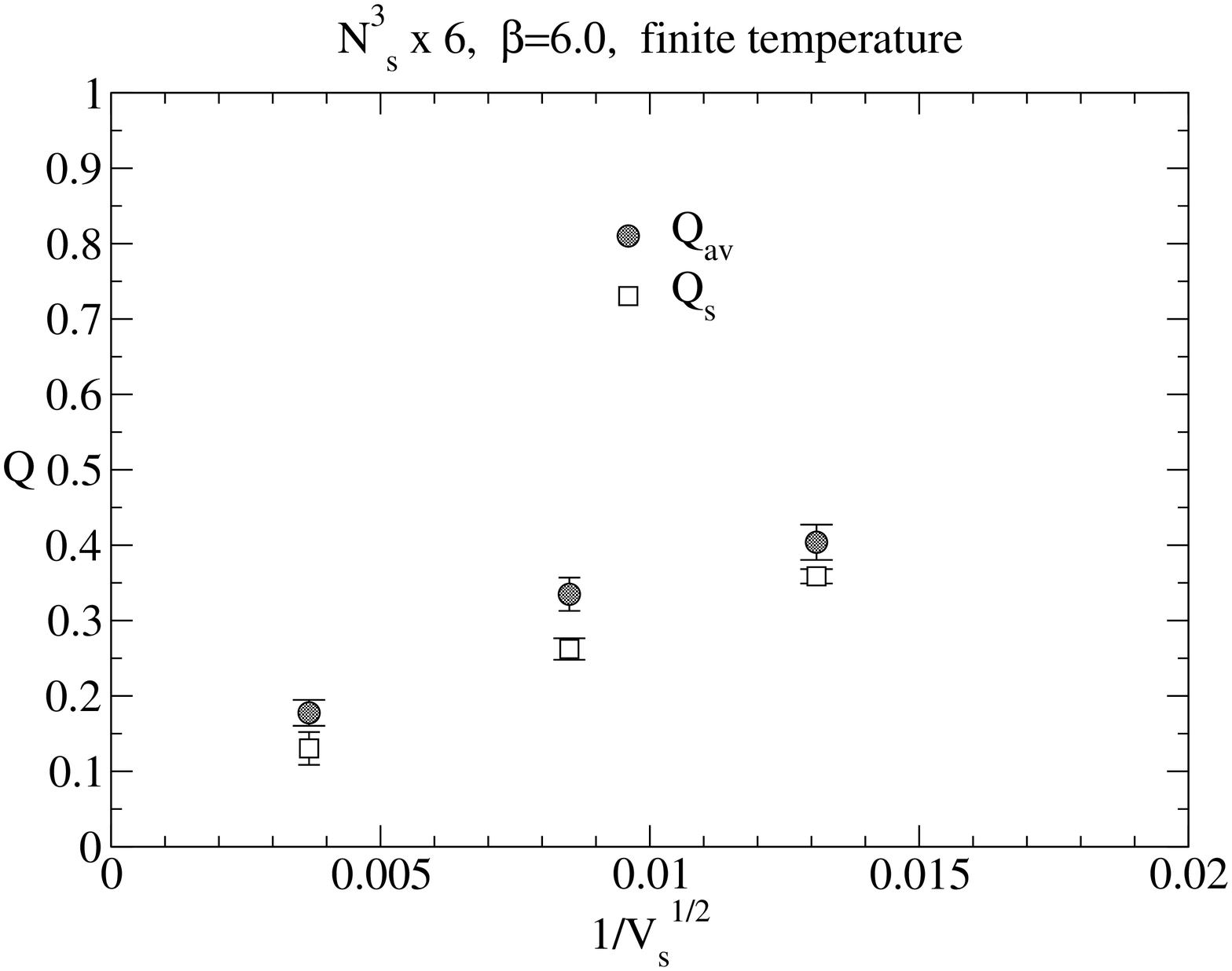}}
\caption{
Volume dependence of $Q$ in the deconfinement phase
at finite temperature, $\T/\T_c = 1.26$.
$Q_s$ and $Q_{av}$ vanish as $V \rightarrow \infty $.
}\label{sqvf}
\end{center}
\end{figure}

We investigate the role of the remnant symmetry in Coulomb gauge. 
We calculate the order parameter $Q_s$ given in Eq. (\ref{qs}), 
and here we also study the color average order parameter
\begin{equation}
Q_{av} = \frac{1}{N_t}
\sum_{t=1}^{N_t} \left<\sqrt{\frac{1}{3}
\mbox{Tr} \tilde{U}(t) \mbox{Tr} {\tilde{U}^{\dagger}(t) }
}\right>. \label{qav}
\end{equation}
The volume dependence of the $Q$ values at zero temperature 
is shown in Fig. \ref{sqv} and the simulation parameters 
are listed in Table \ref{t2}.
It is found that the $Q$ values vanish as $V_s \rightarrow \infty$.
This indicates that the remnant symmetry is unbroken, and the 
system is in the confinement phase.
The $Q$ values in the deconfinement phase are displayed in Fig. \ref{sqvf}.
The simulation parameters are also listed in Table \ref{t2}.
Figure \ref{sqvf} shows that
the $Q$ values at finite temperature, $\T/\T_c \sim 1.26$, 
also go to zero as $V_s \rightarrow \infty $.
It is found that the $Q$ values calculated here indicate
 no qualitative difference 
between the confinement and deconfinement phases
in pure $SU(3)$ gauge theory; 
it is surprising that all the $Q$ values vanish as $V_s \rightarrow \infty$ 
even in the deconfinement phase, but this is a consequence of the fact
that the Coulomb heavy quark potential $V(R,0)$ is always confining.

\begin{table}[h]
\caption{Simulation parameters for the calculation of $Q_{av}$ and $Q_{s}$.
All configuration measurements were
performed every 10 steps.}
\begin{center}
\begin{tabular}{cccc}
\hline
\hline
\multicolumn{2}{c}{$N_s^3\times 32$  } &
\multicolumn{2}{c}{$N_s^3\times 6 $  }\\
\hline
$N_s$ & No. of conf. & $N_s$ & No. of conf. \\
\hline
8 &  100& 18 &  100 \\
18&  100& 24 &  90  \\
24&  100& 42 &  30  \\
32&  30 &    &      \\ 
\hline
\end{tabular} \label{t2}
\end{center}
\end{table}

% \subsubsection{Remnant symmetry at finite temperature}

\subsection{Heavy quark potential in Lorentz gauge}

\begin{figure}[htbp]
\begin{center}
\resizebox{13cm}{!}{\includegraphics{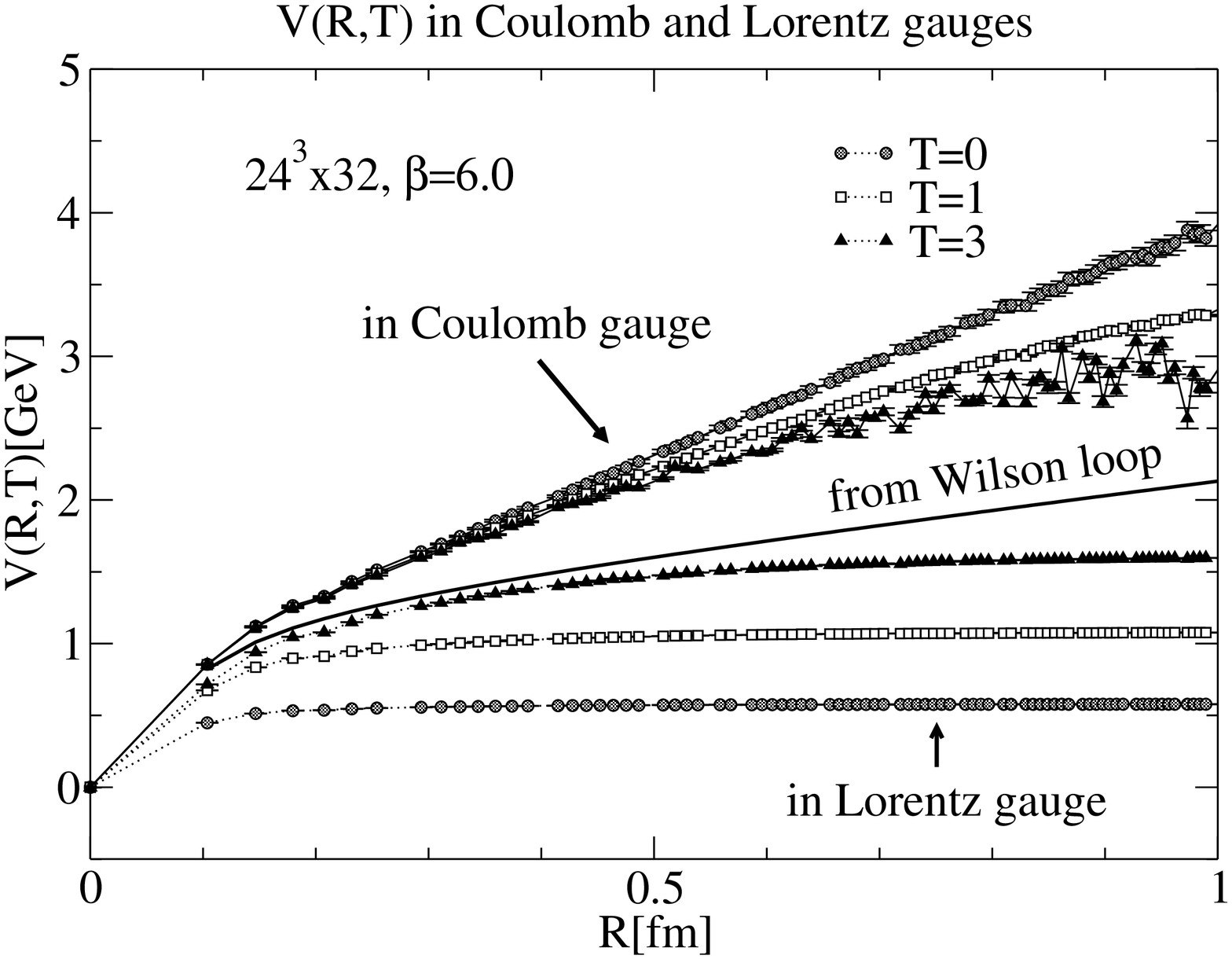}}
\caption{Heavy quark potentials in Coulomb and Lorentz gauges at $\beta=6.0$.
The symbols with error bars represent the numerical data
in Coulomb and Lorentz gauges, and the thick solid curve represents  
the Wilson loop potential at $\beta=6.0$.
}\label{spl}
\end{center}
\end{figure}

Although the argument concerning the Coulomb heavy quark potential
defined by the PPL correlator \cite{Greensite, Greensite2}
is based on use of Coulomb gauge,  
\footnote{Here we use the Lorentz gauge, $\partial_{\mu} A_{\mu}=0$, 
and its realization on the lattice is performed using the procedure
described by Eqs. (2.6) and (2.7), except that the indices
of the link variables are taken as $\mu=1-4$.}
{we carry out the same calculation in Lorentz gauge.}
The results are displayed in Fig. \ref{spl}. 
This calculation was performed on the $24^3 \times 32$
 lattice at $\beta=6.0$, and 100 configurations measured
 every 100 steps are used.

The potential $V(R,0)$ in Lorentz gauge represents
 completely different behavior  
compared with the case of the Coulomb gauge.
It is flat even at large distances.
Moreover, as $T$ increases, the Lorentz heavy quark potential tends to 
approach the usual Wilson loop potential from below, but
at large distances, the Lorentz heavy quark potential
 seems not to be confining.
This strongly suggests that the color confinement mechanism  
is very different between Coulomb and Lorentz gauges. 

\section{Concluding remarks}

We have nonperturbatively studied the long-range behavior of 
the Coulomb heavy quark potential
defined by the partial-length Polyakov line correlators 
in quenched $SU(3)$ lattice gauge simulations.
We confirmed numerically that the Coulomb heavy quark potential, 
corresponding to the color-Coulomb instantaneous part of the time-time
gluon propagator, 
is confining, as suggested in the Coulomb confinement scenario \cite{CCG}.
The $SU(3)$ results obtained in this study
 are qualitatively consistent with those obtained in the $SU(2)$ analysis
 carried out by Greensite et al. \cite{Greensite,Greensite2}.
It is significant that we treated the 
instantaneous part nonperturbatively in the numerical simulations.

The Coulomb heavy quark potential $V(R,0)$ in the confinement phase
rises linearly at large distances. 
Its string tension is several times larger than that obtained from the
usual Wilson loop potential.
As the temporal extension, $T$, of the PPL correlator, increases,
the Coulomb heavy quark potential asymptotically  approaches
the usual Wilson loop potential.
In these simulations, complete agreement is not confirmed.
Note that consistency was found in the
$SU(2)$ lattice gauge simulation \cite{Greensite}.
Furthermore, the result that
the $SU(3)$ Coulomb heavy quark potential indicates
a stronger confining property can be understood from the relation 
$V_{phys}(R) \le V_{coul}(R)$\cite{Zwan}.

The Coulomb string tension can also be estimated
 from lattice calculations of gluon propagators in Coulomb gauge. 
It is reported in Ref. \cite{CZproc} that 
the Coulomb string tension $\sigma_{coul} \sim \sigma_{wilson}$. 
However, in the extensive lattice study
 of Langfeld and Moyaerts \cite{Langfeld},   
the authors concluded that the result
 $\sigma_{coul} \sim (2-3) \sigma_{wilson}$  
cannot be ruled out. This observation is consistent with results 
derived in $SU(2)$ and $SU(3)$ lattice calculations 
with the PPL correlator.  

In $SU(2)$ lattice simulations \cite{Greensite2},
$\beta$ dependence of the Coulomb string tension
scales as a two-loop $\beta$ function.
The Coulomb string tension measured here depends on the QCD coupling 
$\beta=6/g^2$.
This $\beta$ dependence is not explicable here, and it is
therefore desirable to
verify it with numerical simulations in higher $\beta$ regions.

The Coulomb heavy quark potential $V(R,0)$ 
 in the deconfinement phase at $\T/\T_c \sim 1.50$,
 also increases linearly with $R$ at large distances;
 i.e., it is not screened.
However, the Coulomb heavy quark potential
with the finite temporal length in Coulomb gauge
is sufficiently screened, and finally it becomes comparable to 
the screened potentials obtained from the full temporal length Polyakov 
line correlator.
This result may be explained in the Coulomb confinement scenario
\cite{CCG,renorm}. 
$V(R,0)$ corresponds to the instantaneous part, $V_{coul}(R)\delta(t)$,
and therefore $V_{coul}(R)$ does not depend on the time (temperature), 
whereas the non-instantaneous vacuum polarization term causes 
the time (temperature) dependent contribution.

We have also investigated the behavior of an 
 order parameter related to the remnant symmetry in Coulomb gauge.
As the lattice volume approaches infinity, 
the order parameter $Q$ vanishes
in both the confinement and deconfinement phases. 
Thus, this order parameter is not a good parameter
for understanding confinement physics, at least
 in pure $SU(3)$ gauge theory.
This feature is a consequence of the fact that the Coulomb heavy quark
potential is always confining in the confinement and deconfinement phases.

The Coulomb gauge  plays an essential role in the confinement scenario 
discussed in Ref. \cite{CCG}.
We carried out the same calculation for the Lorentz gauge 
and found that the behavior of the Lorentz heavy quark potential
is completely different from the case of Coulomb gauge.
In Lorentz gauge, a different confinement scenario
should be called for.

The color-Coulomb instantaneous part of the time-time gluon propagator
exhibits color confinement.
As discussed in Refs. \cite{CCG} and \cite{renorm}, this expectation may be 
satisfied in a dynamical-quark lattice simulation.
The vacuum polarization causes a quark-pair creation at large quark 
separations, i.e. ``string breaking'', whereas 
the Coulomb linear-rise potential also exists in that case.
If one nonperturbatively extracts the contribution of the vacuum polarization,
it may be easier to see the string breaking.

The color-Coulomb instantaneous part, $V_{coul}(R)\delta(t)$ or $V(R,0)$
on the fixed-time slice, shows the strong linear-rise
behavior in $SU(2)$ and $SU(3)$ gauge theories
at zero and finite temperatures;
this phenomenon is as expected in view of the Coulomb confinement
scenario \cite{CCG,renorm}.
In terms of the Faddeev-Popov (FP) operator $M$,
$V_{coul}(R)$ is given by 
$\langle M^{-1} (-\partial_i^2) M^{-1} \rangle$, 
which has time-independent and long-range properties.  
According to the Gribov picture, 
 gauge configurations for which the eigenvalues of the FP operator are small, 
i.e., regions near the Gribov horizon, 
play an important role in the confinement dynamics \cite{Gribov}. 
Greensite et al. investigated eigenvalue distributions 
of the FP operator
 using the $SU(2)$ lattice gauge simulation \cite{FPeigen}.  

Here we concentrate on the Coulomb gauge. 
It would be interesting to study how the picture changes
 when we employ the other gauge
 by interpolating two gauges, such as
$\sum_{i=1}^3 \partial_i A_i + \lambda \partial_t A_t=0$\cite{IPG}.

In this study, we use the gauge rotation, $e^{i\alpha \partial_i A_i}$, 
which is used in the stochastic gauge fixing\cite{SGF}.
This fixing term is ``attractive" in the Gribov region and 
``repulsive" if one eigenvalue of the Faddeev-Popov operator 
 is negative, 
and therefore the effect of Gribov copy may not be so serious. 
Nevertheless, it is important to study this point.

We investigated the behavior of the color-singlet $q\bar{q}$ 
potential in the present calculation.
 However, an extensive numerical study of the 
color-dependent forces between two quarks may be necessary 
to understand multiquark hadrons.
In Coulomb gauge, $SU(3)$ color-dependent potentials have also been 
calculated in lattice gauge simulations\cite{zeroCDP}.

It is interesting from a phenomenological point of view that 
the linearly rising behavior of the Coulomb potential
exists even in the deconfinement phase at finite temperature.
Therefore, a higher temperature simulation is indispensable and 
it may help to elucidate the behavior of 
 the complex quark-gluon plasma system.

\section{Acknowledgments} 

We would like to thank D. Zwanziger for many helpful discussions
and his continuous encouragement. 
We are also grateful to H. Toki for useful comments.
The simulations were performed on an SX-5 (NEC) vector-parallel computer
at the RCNP of Osaka University. 
We appreciate the support of the RCNP administrators. 
This work is supported by Grants-in-Aid for Scientific Research from 
 Monbu-Kagaku-sho (No. 11440080, No. 12554008 and No. 13135216).
% This work is supported by Grants-in-Aid for Scientific Research from 
% Monbu-Kagaku-sho (Nos. 11440080, 12554008 and 13135216).

\end{document}